\def\be{\begin{eqnarray}}
\def\ee{\end{eqnarray}}
\def\half{\textstyle{1\over 2}}
\def\pis{\Pi\hspace{-0.075in}/}

\documentstyle{article}
\font\tenbf=cmbx10
\font\tenrm=cmr10
\font\tenit=cmti10
\font\elevenbf=cmbx10 scaled\magstep 1
\font\elevenrm=cmr10 scaled\magstep 1
\font\elevenit=cmti10 scaled\magstep 1

\renewenvironment{thebibliography}[1]
 { \elevenrm
   \begin{list}{\arabic{enumi}.}
    {\usecounter{enumi} \setlength{\parsep}{0pt}
     \setlength{\itemsep}{3pt} \settowidth{\labelwidth}{#1.}
     \sloppy
    }}{\end{list}}

\begin{document}
\begin{center}{{\tenbf THE ZERO TENSION LIMIT\\
               \vglue 10pt
               OF STRINGS AND SUPERSTRINGS\\}
\vglue 1.0cm
{\tenrm ULF LINDSTR\"{O}M \\}
\baselineskip=13pt
{\tenit  Department of Physics, University of Stockholm, \\}
\baselineskip=12pt
{\tenit Vanadisv\"{a}gen 9, S-113 46 Stockholm, Sweden\\}
\vglue 0.8cm
{\tenrm ABSTRACT}}
\end{center}
\vglue 0.3cm
{\rightskip=3pc
 \leftskip=3pc
 \tenrm\baselineskip=12pt
 \noindent
The string equivalent of a massless particle  ($m=0$) is the tensionless
string
($T=0$).
The study of such strings is of interest when trying to understand the high
energy limit of
ordinary strings. I discuss the classical $T\to 0$ limit of the bosonic
string,
the spinning
string and the superstring.  A common feature is the appearence of a
space-time
(super-)
conformal symmetry  replacing the world-sheet Weyl invariance. The question of
whether this
symmetry may survive quantization is addressed. A lightcone analysis of the
quantized bosonic
tensionless string leads to severe constraints on the physical states: they
are
space-time
diffeomorphism singlets characterized by their topological properties only.

\vglue 0.6cm}
{\elevenbf\noindent 1. Introduction}
\vglue 0.2cm 
\baselineskip=14pt \elevenrm
Both the point particle and the string have actions of
the type \be
S=T\int{\cal L} \label{act1}
\ee
where $T$ is a dimensionful parameter. In the particle case we have good
reasons
to be
interested in the limit $T\to 0$, since this describes massless particles.
These
we want to study in their own right or as models of particle behaviour at high
energies. Similarily it is interesting to consider strings in the limit $T\to
0$, which  presumably would be a useful description of strings at high
energies
\cite{GROS1}.
The study of such strings turns out to be quite rewarding: as I will mention
at
the end of this talk, they may provide a link to the "topological phase" of
string theory.

This review describes work that I have done in various collaborations (see the
acknowledgement). A
number of other authors have studied tensionless strings
\cite{ZELT}-\cite{BARC}, originating with
the work of Shild \cite{SHIL}.

\vglue 0.6cm {\elevenbf\noindent 2. The Bosonic String}
\vglue 0.4cm

The way to construct the $T\to 0$ limit of the action (\ref{act1}) used in the
point particle
case is to write
\be
S\to \int{\Phi {\cal L}^2+\Phi^{-1}T^2}.\label{act2}
\ee
Integrating out the auxiliary field $\Phi$ one recovers the previous action,
but
in (\ref{act2})
the limit $T\to 0$ is immediate. This procedure has been used to find a
bosonic
tensionless
string too \cite{KARL}, but it is only for the particle that this procedure
has
any relation
to the world-volume geometry. (There $\Phi$ is identified with the "einbein"
$e$.) It has proven
useful to keep a relation to the geometry and I will describe a procedure for
doing so. To keep
things general I start from the action for a $p$-brane \footnote{This section
is
based on
material derived in collaboration with R.v.Unge}
\be
S=T\int{d^{p+1}\xi \sqrt{-\gamma}}\label{mact}
\ee
where $\gamma \equiv det\gamma_{ab}$ and the line element is
\be
ds^2=dX^\mu dX^\nu \eta_{\mu \nu}=\partial_aX^\mu \partial_bX^\nu \eta_{\mu
\nu}d\xi^ad\xi^b\equiv \gamma_{ab} d\xi^ad\xi^b.
\ee
($\gamma_{ab}$ is the induced metric on the world-volume.) We pass to the
Hamiltonian
formulation via the canonical momenta derived from (\ref{mact}):
\be
P_\mu =T\sqrt{-\gamma}\gamma^{a0}\partial X_\mu .
\ee
They obey the constraints
\be
P^2+T^2\gamma\gamma^{00}=0,\qquad P_\mu \partial_iX^\mu =0,\quad i=1,...,p.
\ee
The Hamiltonian is just these constraints multiplied by Lagrange multipliers
that I call
$\lambda$ and $\rho^i$,
\be
{\cal H}=\lambda
\left({P^2+T^2\gamma\gamma^{00}}\right)+\rho^iP\cdot\partial_iX,
\ee
where $\cdot$ denotes contraction using the space-time metric $\eta$.
The corresponding phase space action is
\be
S^{PS}=\int{d^{p+1}\xi\left\{{P\cdot \dot{X}-\lambda
\left({P^2+T^2\gamma\gamma^{00}}\right)-\rho^iP\cdot\partial_iX }\right\}}.
\ee
To obtain a new configuration space action we integrate out the momenta
$P_\mu$
to find
\be
S_2^{CS}=\half\int{
d^{p+1}\xi\left({{1\over {2\lambda}}}\right)\left\{{\dot
X^2-2\rho^i\dot X\cdot\partial_iX+\rho^i\rho^j\partial_iX\cdot\partial_jX
-4\lambda^2T^2\gamma\gamma^{00}}\right\}}.\label{CSACT1}
\ee

As an aside we note that integrating out also the $\rho^i$'s we recover the
form
(\ref{act2})
after a suitable identification between $\lambda$ and $\Phi$.

For $p=1$ we have only one $\rho^i$ and may now identify the Lagrange
multipliers with
componenets of the $D=2$ metric directly, thus arriving at the
Brink-Howe-DiVecchia-Deser-Zumino
\cite{BRIN},\cite{DESE} form of the bosonic string. For general $p$ we have to
go through one
more step.  We rewrite (\ref{CSACT1}) using a $p$-dimensional auxiliary metric
$G_{ij}$ and a
rank $1$ matrix $h^{ab}$:
\be
S^{CS}_1=\half\int{d^{p+1}\xi\left\{{{{h^{ab}\gamma_{ab}}\over{2\lambda}}-2\lambda
T^2G(p-1)
+2\lambda GG^{ij}\gamma_{ij}}\right\}}\label{CSACT2}
\ee
where
\be
h^{ab}=
\left( \matrix{1\quad\hfill -\rho  ^i\cr
 -\rho  ^i\quad\hfill\rho ^i\rho ^j\cr}
\right)\label{hdef}
\ee
and $G\equiv detG_{ij}$. The equivalence between (\ref{CSACT1}) and
(\ref{CSACT2}) is seen by
integrating out $G_{ij}$. If we now keep $T\not =0$, we may make the
identification
\be
g^{ab}=
{1\over 4}T^{-2}\lambda^{-2}G^{-1}\left( \matrix{-1\quad\hfill \rho  ^i\cr
 \rho  ^i\quad\hfill-\rho ^i\rho ^j+4\lambda^2T^2GG^{ij}\cr}
\right)\label{gdef}
\ee
to find the usual $p$-brane action \cite{HOWE1} in terms of the auxiliary
world-volume metric
$g_{ab}$:
\be
S_g=-{T\over 2}\int{d^{p+1}\xi
\sqrt{-g}\left\{{g^{ab}\gamma_{ab}-(p-1)}\right\}}.
\ee

In the limit $T\to 0$ we instead make the identification
\be
V^a\equiv {1\over{\sqrt{2}\lambda}}\left({1,-\rho^i}\right).
\ee
This yields the action for the zero tension theory. For the string it reads
\be
S_V=\int{d^2\xi V^aV^b\gamma_{ab}}\label{Vstring}
\ee
with $V^a$ a world-sheet vector density. This form of the action has proven
very
useful. For
example it is readily supersymmetrized to give the zero tension limit of the
superstring and
of the spinning string. The "geometrical" structure also made it possible to
write down
a number of models with world-sheet supersymmetry by constructing a new type
od
$D=2$
"degenerate" supergravity \cite{LIND1}.

In taking the $T\to 0$ limit {\it the Weyl-invariance of the string theory has
been replaced by space-time conformal invariance}. This is clear from the fact
that a space-time conformal transformation will scale the induced metric, a
scaling that may be absorbed by a compensating scaling of $V^a$. This was not
possible for the tensionful action
\be
S_g=T\int{d^2\xi\sqrt{-g}g^{ab}\gamma_{ab}},
\ee
since $\sqrt{-g}g^{ab}\gamma_{ab}$ is invariant by itself under scalings.
That there should be such a symmetry is natural from the description of the
theory in the $2D$-diffeomorphism gauge $V^a=(v,0)$ where the field equations
read
\be
\ddot{X}^\mu (\xi)=0,\qquad\dot{X}^2(\xi)=\dot{X}^\mu (\xi)X'_\mu (\xi)=0,
\ee
i.e., for each $\sigma$ we have a massless particle moving on a null
trajectory
(and satisfying an orthogonality constraint). Conformal symmetry is precisely
the symmetry that preserves the light-cone (causal) structure.

There are two topics based on the classical formulation (\ref{Vstring}) of the
tensionless string that I want to introduce: The question of
supersymmetrization and the quantization of the bosonic string.

\vglue 0.6cm {\elevenbf\noindent 3. The Superstring}
\vglue 0.4cm

The zero tension limit of the superstring \cite{LIND2} is achieved by a
(space-time) supersymmetrization of the bosonic model (\ref{Vstring}):
\be
\partial X^\mu \to \Pi^\mu_a \equiv \partial X^\mu -i\bar{\theta}\Gamma^\mu
\partial_a\theta\label{pdef}
\ee
with $\theta (\xi)$ are Majorana (or Weyl) space-time spinors. The action
becomes
\be
S_V\to \int{d^2\xi V^aV^b\Pi^\mu_a\Pi_{b\mu}}
\ee
The symmetries of this action are (global) Space-time supersymmetry,
(in certain dimensions it is even superconformally invariant),
and (local) Siegel-invariance.

The global supersymmetry is evident from the
definition (\ref{pdef}). Superconformal invariance follows as in the bosonic
case, since a superconformal transformation will scale the super-line element
\be
ds^2_S=\eta_{\mu \nu}\Pi^\mu_a\Pi^\nu_b d\xi^ad\xi^b.
\ee
The superconformal group exists in $D=2-6$. The
Siegel invariance is given by
\be
\delta^\kappa \theta=i\pis_a\kappa^a,\qquad \delta^\kappa
X^\mu=i\bar\theta\Gamma^\mu \delta^\kappa \theta ,\label{kappa1}
\ee
where $\kappa^a$ is a world-sheet vector space-time spinor. For $T\not =0$ one
has to add a Wess-Zumino term to the supersymmetrization of the
bosonic action to achieve $\kappa$-invariance. This is not the necessary in
the
$T=0$ case at hand. With
\be
\kappa^a\equiv V^a\kappa,
\ee
where $\kappa$ is a density of opposite weight to $V^a$, and
\be
\delta_\kappa V^a=2V^aV^b(\partial_b\bar\theta)\kappa,\label{kappa2}
\ee
the model is Siegel-invariant as it stands. (In fact,the WZW-term is
separately
invariant under the
transformations (\ref{kappa1}-\ref{kappa2}) in $D=2$ $mod8$.) On shell the
$\kappa$-dependence is
through the combination $V^a\pis _a\kappa $ only, and $V^a\pis _a$ is
nilpotent
there. Hence
$\kappa$ carries half the degrees of freedom of a spinor in complete analogy
to
the $T\not =0$ case.
For closure of the Siegel symmetry one finds a version of the usual local
bosonic symmetry \be
\delta_\lambda V^a=0,\quad \delta_\lambda \theta =\lambda
V^a\partial_a\theta,\quad \delta_\lambda
X^\mu =i\bar\theta \Gamma^\mu\delta_\lambda \theta. \ee
In the diffeomorphism gauge $V^a=(v,0)$ we again find that the equations of
motion describe, for each $\sigma$, a massless {\it {super}}particle
moving on a superspace equivalent of a null hypersurface:
\be
\dot{\Pi}^\mu_0=0,\quad\bar\theta\pis_0=0,\quad
(\Pi_0)^2=0,\quad\Pi_0\cdot\Pi_1=0.
\ee

\vglue 0.6cm {\elevenbf\noindent 4. The Spinning String}
\vglue 0.4cm

The zero tension limit of the spinning string can likewise be constructed
starting from the bosonic action (\ref{Vstring}) \cite{LIND3}. A world-sheet
supersymmetrzation leads to
\be
S=\int{d^2\xi \left\{{(V^a\partial_aX^\mu+i\Psi^\mu\chi )(V^b\partial_bX_\mu
+i\Psi_\mu\chi)+i\Psi^\mu V^a\partial_a\Psi_\mu}\right\}},\label{Spinact}
\ee
where $\chi$ is the fermionic partner of $V^a$ and $\Psi^\mu$ is that of
$X^\mu$.(See further \cite{LIND3}). The world-sheet "spinors" are really
Grassmann numbers here. The world-sheet supersymmetry transformations that
leave (\ref{Spinact}) invariant are
\be
\delta X^\mu &=&i\varepsilon \Psi^\mu, \qquad \delta\Psi ^\mu =-\varepsilon
\partial X^\mu -\half i\varepsilon (\Psi^\mu \chi)\cr
\delta V^a&=&iV^a(\varepsilon \chi),\qquad \delta \chi =\nabla \varepsilon ,
\ee
where $\varepsilon$ is a spinor density, $\partial \equiv V^a\partial_a$ and
$\nabla
\equiv V^a\nabla_a$. The covariant derivative involves a connection about
which it is sufficient to assume $\nabla_aV^a=0$, which thus is the "metricity
condition" of our theory.

In the gauge  $V^a=(v,0)$, we find that the model describes the motion of a
massless spinning particle for each $\sigma$.

The action (\ref{Spinact}) of this theory closely resembles that of a massless
spinning particle as studied by Howe, Pernati, Pernici and Townsend
\cite{HOWE}.
Just like that model it can be extended to abitrary $N$ number of
supersymmeties
and carry a gauged $O(N)$ symmetry. The $(1,1)$ model is the zero tension
limit
of the spinning string.

Further, introducing a degenerate local superspace supergravity theory
corresponding to the $(V^a,\chi)$ multiplet \cite{LIND1}, the action
(\ref{Spinact}) takes the form
\be
S=\int{d^2\xi d\theta\nabla {\cal X}^\mu\nabla^2{\cal X}_\mu},
\ee
where the superfields and superspace covariant derivatives are
\be
\nabla &=& E\partial_\theta +E^a\partial_a,\cr
E^a \mid &=&V^a,\quad \nabla E\mid=\left({\textstyle{2\over
3}}\right)\chi,\qquad {\cal
X}^\mu=X^\mu +\theta \Psi^\mu .
\ee
The superspace formulation leads to the introduction of a number of new
models.
E.g., the zero tension limit of the $(2,2)$ string is introduced via a
complexification of the ${\cal X}^\mu$ field.

\vglue 0.6cm {\elevenbf\noindent 5. The Quantum Theory}
\vglue 0.4cm

To investigate the quantum properties of the bosonic theory one may of course
proceed in
several ways, as for the tensionful string. The only avenue which is not open
is
to demand
Weyl-invariance of the quantum theory.

Both the hamiltonian and the
lagrangian BRST versions with diffeomorphism ghosts lead to nilpotent BRST
charges independent
of the dimension of space-time. Likewise for closure of the Lorentz algebra in
the light-cone
gauge: no obstruction, no critical dimension.

However, as was mentioned previously, the Weyl-invariance of the $T\not =0$
string is
replaced by global conformal symmetry of the ambient space-time. This brings
up
the question
of survival of this symmetry in the quantum theory, a question addressed in
\cite{LIND4}.

After going to light-cone gauge and solving the constraints we are left with
highly
non-linear expressions for the conformal generators in terms of the
transversal
coordinates $X^i(\sigma)$ and momenta $\Pi_i(\sigma)$. Skipping the details of
how to ensure
hermiticity of the operators, choosing a reference operator ordering,
regularization et.c.,
canonical quantization is achieved by prescribing the operator commutation
relations
\be
\left[{X^i(\sigma ),\Pi^j(\tilde \sigma)}\right]=i\delta^{ij}\delta(\sigma
-\tilde \sigma),
\qquad\left[{X^-,\pi^+}\right]=-i,
\ee
where $+$ and $-$ refer to the two lightlike directions.
We then find a systematic and controlled way of calculating the conformal
algebra. The
Lorentz subalgebra closes indeed. However, leaving that subalgebra we
immediately encounter
obstructions: operator anomalies. E.g., the commutator between the generators
of
special
conformal and Lorentz transformations, $K$ and $M$, should close to $K$.
Instead
we find
\be
\left[{
K^i,M^{j-}
}\right] -\delta^{ij}K^-\propto
\int{
d\sigma
\left({
{1\over{\varepsilon\pi^+}}
}\right)
\left\{{
X^i\Pi^j-x^i\pi^j
}\right\}
}\equiv
\left({
{1\over{\varepsilon\pi^+}}
}\right)L^{ij}\not = 0.
\ee
Lower case letters denote zero-modes, i.e. $\sigma$ independent pieces of the
operators.
(Note, in passing, that for the particle theory the zero-modes are the whole
story, and the
obstructions vanish. The conformal symmetry is a good quantum symmetry fort
the
massless
particle.)

The route we choose from here is to impose
\be
L^{ij}|{PHYS}>=0\label{phys}
\ee
and see if we can carry on this procedure to find the state space of the
quantum
theory. We
now have to commute $L^{ij}$ with the conformal generators. Eventually this
procedure will
terminate and leave us with some algebra. Invoking a result by Ogievetsky
\cite{OGIE}, this
closure is the algebra of general coordinate transformations. \footnote{In
fact,
the closure yields
only the {\it {analytic}} diffeomorphisms. I thank M.B. Green and G.
Papandopolous for this comment.}
Thus, for the space-time conformal group to survive as a symmetry in the
quantized theory the physical
states must be singlets under general coordinate transformations! The states
of
the theory should hence
correspond to equivalence classes of string configurations differing only in
their {\it topological
properties}.

\vglue 0.6cm {\elevenbf\noindent 6. Discussion}
\vglue 0.4cm

Admittedly, the route taking in imposing the constraints (\ref{phys}) is
unorthodox. Normally one
would perhaps have concluded that there are anomalies and that the space-time
symmetery is not
a good quantum symmetry. However, our goal was precisely to investigate under
what circumstances the
conformal symmetry survives as a quantum symmetry. One interpretation is then
that we modify the
theory so as to include the constraints and that this {\it {defines}} our
quantum theory.

One should
also bear in mind that we are trying to find the large symmetry that Gross
finds
indicatins of in
studying high energy scattering of strings. It would be nice to be able to say
something conclusive
about that topic, but all we can say is that our results do not contradict
those
of Gross. The
selection rules he finds say that in the $T\to 0$ limit amlitudes are give by
polarizations (spin) in
the scattering plane, i.e., the plane defined by the relative momenta. Other
polarization directions
do not affect the amplitudes. The constraint (\ref{phys}) imply that no spin
is
allowed for a single
tensionless string, but on the other hand there is no relative momentum in
this
single string Hilbert
space.

Finally I emphasize that we have only found restrictions on the Hilbert space,
we have no explicit
construction of the spectrum. The result seems to tie in nicely with the ideas
of unbroken general
covariance and few short-distance degrees of freedom, though.

\vglue 0.6cm {\elevenbf\noindent  Acknowledgements:} It is a great pleasure to
thank my
collaborators on the various papers upon which this report is based: Jan
Isberg,
Anders
Karlhede, Martin Ro\v cek, Bo Sundborg, Georgios Theodoridis and Rikard von
Unge.

\vglue 0.6cm

\end{document}